# Deterministic Service Function Chaining over Beyond 5G Edge Fabric

Hao Yu∗, Tarik Taleb∗† and Jiawei Zhang‡ ∗Aalto
University, Espoo, Finland.
†Oulu University, Oulu, Finland.
‡Beijing University of Posts and Telecommunications, Beijing, China.

*Abstract*—Along with the increasing demand for latencysensitive services and applications, Deterministic Network (DetNet) concept has been recently proposed to investigate deterministic latency assurance for services featured with bounded latency requirements in 5G edge networks. The Network Function Virtualization (NFV) technology enables Internet Service Providers (ISPs) to flexibly place Virtual Network Functions
(VNFs) achieving performance and cost benefits. Then, Service Function Chains (SFC) are formed by steering traffic through a series of VNF instances in a predefined order. Moreover, the required network resources and placement of VNF instances along SFC should be optimized to meet the deterministic latency requirements. Therefore, it is significant for ISPs to determine an optimal SFC deployment strategy to ensure network performance while improving the network revenue. In this paper, we jointly investigate the resource allocation and SFC placement in 5G edge networks for deterministic latency assurance. We formulate this problem as a mathematic programming model with the objective of maximizing the overall network profit for ISP. Furthermore, a novel Deterministic SFC deployment (Det-SFCD) algorithm is proposed to efficiently embed SFC requests with deterministic latency assurance. The performance evaluation results show that the proposed algorithm can provide better performance in terms of SFC request acceptance rate, network cost reduction, and network resource efficiency compared with benchmark strategy.

## I. INTRODUCTION

5G or beyond 5G are supposed to support various emerging vertical industries, e.g., automotive, intelligent manufacturing, which impose unique requirements regarding flexibility, reliability and especially strict QoS guarantee. Deterministic end-to-end latency bound and lower jitter should be provided by provisioning appropriate network resources to the critical services. To this end, Deterministic Network (DetNet) [1] concept has been recently proposed by the IETF DetNet Working Group to initiate the study of deterministic data paths for real-time applications (e.g., industrial automation) with lower packet delay variation and bounded end-to-end latency [2]. We extend this concept into 5G edge networks which facilitate flexible and low-latency network services. With the advent of Network Function Virtualization (NFV) [3] [4] which decouples traditional network functions (e.g., firewalls, load balancers) from dedicated hardware to Virtual Network Functions (VNFs) running on commercial servers for low expense and efficient management, network services can be customized to satisfy the various service requirements in terms of bandwidth, latency and reliability by Internet Service Provides (ISPs). Moreover, to meet the ultra low latency requirements, VNFs are pushed to the edge of the networks for further latency reduction. When dealing with the service requests from 5G mobile users, it is often needed to steer the traffic from the cell sites (CSs) to edge servers by traversing the VNFs concatenated in a specified order, which is defined as Service Function Chain (SFC) [5], [6], [7].

Despite such advantages of applying NFV in 5G edge environments, some challenges still remain for SFC placement and resource allocation at network edge to ensure the deterministic performance and high efficiency. First, due to the finite physical resources (e.g., bandwidth, memory and CPU) in the network, the optimal resource allocation and path selection should be performed optimally to achieve higher resource efficiency and to avoid resource bottleneck. Second, how to ensure that the service latency is bounded within a deterministic scope also remains a problem yet to be solved [8]. It is desirable to design a joint resource allocation and SFC placement algorithm for deterministic latency assurance due to the above challenges. In this paper, we formulate the problem of joint resource allocation and SFC placement in the 5G edge environment. Note that, we will employ NFV concepts in open and smart radio access network (O-RAN) [9], the baseband processing units in the new O-RAN architecture can be implemented as RAN VNFs instantiated at the edge servers. The deployment of VNF instances needs to consume network resources which will incur the increasing Capital Expense (CAPEX), so it is beneficial to increase the revenue of ISPs by optimizing the resource allocation and placement of SFC.

The SFC deployment and resource management has drawn much research attention. The authors in [10] studied the joint SFC deployment and resource management problem (JSDRM) in heterogeneous edge environments to minimize the total system latency and proposed a scheme based on a game model to jointly deploy SFCs and manage resources. The authors in [11] [12] also investigated the resource re-allocation of SFC strategy due to different mobility patterns. Different from the existing work, aiming at minimizing the service latency and network cost in the literature, deterministic latency assurance [2] [13] is more crucial for the packets of the time-critical flows.

In this paper, we will study the SFC deployment scheme with the objective of achieving deterministic latency assurance while maximizing the profit of ISPs. Note that, minimizing service latency is not the objective of this paper, we try to ensure that the experienced latency is equal to the service latency requirement for a single SFC. Based on this, we maximize the overall profits by proposing an optimal SFC deployment strategy.

The remainder of this paper is organized as follows. Section II presents the system model. Section III formulates the joint resource allocation and SFC placement for deterministic latency assurance problem and our proposed algorithm is presented in Section IV. The performance evaluation results are discussed in Section V. Section VI concludes this paper.

## II. SYSTEM MODEL

We consider a 5G edge network architecture comprised of edge nodes and cell sites forming a multi-cell coverage area for mobile users. We denote by $G = (N,E)$ the physical networks consisting of $N$ physical edge nodes. We use K to denote the set of SFCs. Each SFC $k \in K$ is defined as a vector $\{s_k, d_k, V_k, L_k, \lambda_k, L_k\}$ to dictate the properties of this SFC. $s_k$ and $d_k$ correspond to the source and destination nodes of SFC $k$. Since each SFC $k \in K$ consists of a given number of ordered VNFs, we use $V_k = \{v_{k,1}, v_{k,2},...,v_{k,I_k}\}$ to denote the set of VNFs in SFC $k$, where $I_k$ is the number of VNFs and $v_{k,i}$ is the $i^{th}$ VNF in SFC $k$. We denote the link between the $i^{th}$ VNF and the $(i+1)^{th}$ VNF in SFC $k$ as $l_{k,i}$. Thus, $L_k$ is the set of virtual links of SFC $k$, i.e., $L_k = \{l_{k,0}, l_{k,1},...,l_{k,I_k}\}$. Each SFC supports one service associated with a bit rate $\lambda_k$ which is an aggregated bit rate by multiple users belonging to this SFC $k$. $L_k$ is the end-to-end latency requirement.

## III. DETERMINISTIC RESOURCE ALLOCATION AND SFC PLACEMENT PROBLEM

### A. Problem Description

In a NFV-enabled 5G edge network, ISP should make optimal planning for SFC request deployment to maximize its profit while ensuring the deterministic performance requirements of SFC requests. The problem can be described as: given: a physical network topology $G = (N,E)$ and a set of SFC requests K, for each SFC request, determine: 1) how to allocate computation and bandwidth resources for corresponding NFVs and traffic, 2) how to route the SFC along an appropriate path and where to place VNF instances, 3) maximize: the profit of ISP from deploying SFC requests, at the same time, 4) ensure: deterministic end-to-end latency.

### B. Problem Formulation

The SFC deployment basically consists of VNF instance placement and resource allocation for VNF instances. These two parts are usually interactive with each other and should be jointly considered.

*1) VNF instance placement:* we denote by binary variable $x_{k,i,n}$ the placement of NFV $v_{k,i} \in V$, $x_{k,i,n} = 1$ iff VNF $v_{k,i}$ is placed in edge node $n \in N$, otherwise, $x_{k,i,n} = 0$. Based on the observation that one edge node can instantiate multiple VNF instances and one VNF instance can only run on top of an edge node, the placement constraint of VNF instance $v$ can be given as follows:

$$\sum_{n \in N} x_{k,i,n} = 1, \forall k \in K, \forall i \in V_k \quad (1)$$

Since the computing capability of edge node is shared by all NFV instances that are placed on it, the total CPU and memory capability allocated to VNF instances can not exceed the total capability of the edge node, then we have,

$$\sum_{k \in K, i \in V_k} x_{k,i,n} \cdot \pi_{k,i,r} \leq C_{n,r}, \forall n \in N, r \in R \quad (2)$$

where we define $\pi_{k,i,r}$ to indicate the amount of processing resources of type $r \in R$ allocated to VNF $i$ of SFC $k$. $C_{n,r}$ represents the resource capability of type $r$ in edge node $n$. The unit of CPU resource allocation is the number of CPU cores, and the unit of memory resource allocation is GB.

*2) Traffic routing:* We define the variable $y_{k,i,n,m}$ to denote that routing decision. If the traffic between VNF $i$ and VNF $i+1$ traverses the physical edge $(n,m)$, $y_{k,i,n,m} = 1$, otherwise, $y_{k,i,n,m} = 0$. If $(n,m) \in E$ is traversed by $l_{k,i} \in L_k$, $n, m \in V$ must be traversed as well. Then the constraint must be ensured as:

$$y_{k,i,n} y_{k,i,m} = 1 \text{ if } y_{k,i,n,m} = 1 \quad (3)$$

Furthermore, we need to guarantee that the virtual links on the path to embed SFC request $k$ are connected head-to-tail as:

$$\sum_{\substack{n \in N \\ i \in [0, I_k]}} (y_{k,i,n,m} - y_{k,i,m,n}) = \begin{cases} 1, & \text{if } m = s_k \\ -1, & \text{if } m = d_k \\ 0 & \text{otherwise} \end{cases} \quad (4) \quad (5)$$

If VNF $i$ of SFC $k$ is instantiated on edge node $n$, it must ensure to be traversed as:

$\forall k \in K, i \in V_k, n \in N$

$$x_{k,i,n} \leq y_{k,i,n} \quad (6)$$

Also, since the bandwidth resources of physical edge $(n,m)$ are shared by the virtual links that are mapped on it, the total bandwidth consumed by these virtual links can not exceed the total bandwidth resource of physical edge $(n,m)$. Firstly, we define the following real variable $\eta_{k,i}$, to denote the amount of bandwidth resource allocated to the virtual link between VNF $i$ and VNF $i+1$ of SFC $k$. Then, we add the following constraints. Constraints (7) ensures that the sum of bandwidth resource allocated to virtual links can not exceed the total bandwidth $B_{n,m}$ of physical edge $(n,m)$.

$$\sum_{k\in K, i\in[0,l_k]} y_{k,i,n,m} \times \eta_{k,i} \leq B_{n,m}, \forall (n,m) \in E \quad (7)$$

*C. Deterministic Latency*

For a SFC to be deployed, latency will be incurred by data processing in edge nodes and data transmission in physical edges accordingly, i.e., processing latency and communication latency. The service latency is determined by both resource requirements of SFC and the amount of resources allocated to it.

*1) Processing Latency for VNFs:* for RAN part, the Layer 1 processing is done per assigned Resource Block (RBs) and is mainly dependent on channel condition. The condition in the channel dictates the appropriate coding rate and modulation for the data to be transmitted successfully which leads to different computational demand on Layer 1. Thus, the computational complexity of Layer 1 functions depends on the amount of RBs assigned and Modulation and Coding Scheme (MCS). However, the higher-layer RAN VNFs (i.e., Layer 3 and Layer 2) and other common VNFs (e.g., core network functions) processing are user-load dependent and the computational capacities depend on the aggregated users' data rate. Thus, the total computational demand including RAN protocol stack and other function units for SFC $k$ is given as:

$$c_k = \theta_1 N_k \sum_{j=0}^{2} a_j (i_{MCS,k})^j + \theta_2 \lambda_k \quad (8)$$

where $\lambda_k$ is the aggregated data rate of users, $N_k$ denotes the number of aggregated resource blocks (RBs) allocated to the users in SFC $k$. $a_j$ is the Layer 1 computational resource model-specific constant. $i_{MCS,k}$ is the indices of the MCSs of SFC $k$ as defined in 3GPP TS 38.214 [14]. For the sake of simplicity, we assume that all users within SFC $k$ are assigned with the same MCS indices. $\theta_1$ and $\theta_2$ are the scaling factors of the Layer 1 and other function units in terms of computational resource models.

VNF instances are usually instantiated by associating with a certain combination of resources (e.g., CPU, RAM and storage) and the Layer 1 RAN VNF processing latency can be calculated as the function of CPU frequency allocated if the number of RBs and MCS indices are constant. Given the amount of CPU frequency $\pi_{k,1,1}$ allocated to Layer 1 RAN VNF of SFC $k$, if $r = 1$ means CPU resources and $i = 1$ denotes Layer 1 RAN VNF, the processing time of Layer 1 RAN VNF $v_{k,1,1}$ is formulated as follows according to [15]:

$$\tau_{k,1}^p = \frac{N_k}{\pi_{k,1,1}^2} \sum_{i=0}^{2} a_j (i_{MCS,k})^j \quad (9)$$

Let $\rho_i$ be the amount of CPU resource for one information bit of VNF $i$. The processing latency of the other VNFs $v_{k,i}$ in the SFC $k$ is given as [16]:

$$\tau_{k,i}^p = \frac{\rho_i \lambda_k}{\pi_{k,i,1}} \quad (10)$$

Thus, the total processing latency for VNFs of SFC $k$ is:

$$l_k^p = \sum_{i=1}^{I_k} \tau_{k,i}^p \quad (11)$$

*2) Communication Latency in Virtual Links:* the communication latency of each SFC consists of the propagation latency and transmission latency. Based on the study in literature [17], the communication latency of SFC $k$ on the virtual link $l_{k,i}, i \in \{0,...,I_k\}$ can be formulated as:

$$\tau_{k,i}^c = d_{k,i}^{prop} + d_{k,i}^{trans} \quad (12)$$

Hereby, the first term of Equation (12) indicates the propagation latency on physical edges that virtual link $l_{k,i}$ traverses by, which is computed by the ratio of the length of path to the propagation speed of signals in that medium. The second term indicates the transmission latency, which is calculated by dividing the average size of transmitted packets with the bandwidth capacity allocated to this virtual link:

$$d_{k,i}^{trans} = \frac{b_k}{\eta_{k,i}} \quad (13)$$

Then the communication latency of SFC $k$ is formulated as:

$$l_k^c = \sum_{i=0}^{I_k} \tau_{k,i}^c \quad (14)$$

Finally, considering the deterministic E2E latency requirement $L_k$ of SFC $k$, the E2E latency constraint is given as:

$$l_k^p + l_k^c = L_k \quad (15)$$

*D. Profit Model*

*1) Cost Model:* the cost of SFC $k$ can be defined similar to the one in [18], as follows:

$$C_k = \sum_{i=1}^{I_k} \sum_{r \in R} \alpha_r \pi_{k,i,r} + \sum_{i=0}^{I_k} \beta \eta_{k,i} \quad (16)$$

where $\alpha_r, r \in R$ denotes the cost factor of allocating one resource unit of $r$-type and $\beta$ represents the cost factor of allocating one bandwidth unit.

*2) Revenue Model:* we define the revenue of each SFC $k \in K$ as follows:

$$R_k = \delta \lambda_k + \omega / L_k \quad (17)$$

As the E2E latency requirement $L_k$ can be seen as the most important QoS indicator for the safety-critical services and the performance assurance that it can provide to users, we take into account $L_k$ as the part of revenue that SFC $k$ can earn.

Services with stricter latency constraints will cost more network resources, thus result in more revenues. Besides E2E latency, data rate $\lambda_k$ should also be considered as another QoS indicator by which ISP can charge users, since more network resources (processing and bandwidth resource) will be consumed to provision a service with higher data rate than lower data rate while ensuring the same E2E latency requirement $L_k$. Thus, for each SFC $k$, the overall profit of SFC $k$ is formulated as:

$$P_k = R_k - C_k \qquad (18)$$

*3) The total profit of the system:* the total profit of the system, denoted by $P$, is formulated by the summation of profits of all SFCs deployed as follows:

$$P = \sum_{k \in K} P_k \qquad (19)$$

*4) Problem Formulation:* the joint SFC routing and resource allocation in this paper problem we target is formulated as an optimization problem which maximizes the overall profits of the system:

$$\max P \quad (20) \quad s.t. (1-7)(15) \quad (21)$$

The fact that VNF placement is a NP-hard problem, which has been proved in [19], and Equation (9) is quadratic makes this problem nontrivial to solve. Instead of finding exact numerical solutions by analytical method, which is extremely time-consuming, we propose a novel Det-SFCD algorithm to solve this problem within acceptable timescales.

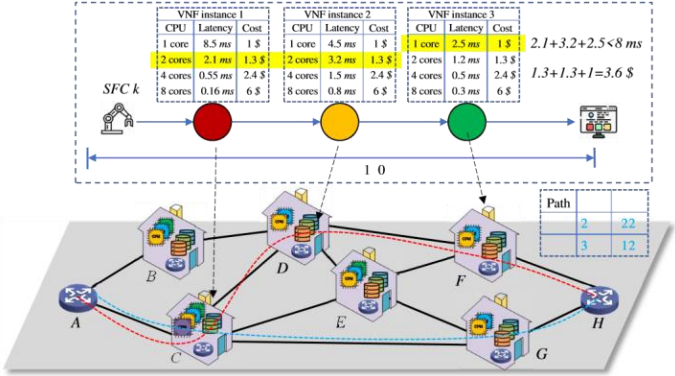

Fig. 1: Joint Resource Allocation and SFC Deployment for Deterministic Latency Assurance.

## IV. DET-SFCD ALGORITHM

In this section, we propose a deterministic SFC deployment (Det-SFCD) algorithm to efficiently deploy SFCs by creating appropriate VNF instances for running the service and deciding an optimal routing path. First, we need to construct a weighted graph where the deployments cost are defined as the weight. Second, based on the resource cost defined in weighted graph, we derive the optimal path with the least deployment cost for SFC deployment. Third, we determine the optimal resource allocation scheme for VNF instances of SFC with minimal resource cost on the selected path while ensuring the deterministic latency.

To obtain optimal path for SFC deployment, we define a metric called "deployment cost" which is considered to balance the network load and reduce resource bottleneck. We use $c^{cpu}_{k,n}$, $c^{mem}_{k,n}$ and $c^{bw}_{k,n,m}$ to represent the deployment costs of CPU on node $n$, memory on node $n$ and bandwidth on link $(n,m)$ when deploying SFC $K$, respectively. Each of them is defined as:

$$c^{cpu}_{k,n} = \frac{1}{r^{cpu}_{k,n}}, c^{mem}_{k,n} = \frac{1}{r^{mem}_{k,n}}, c^{bw}_{k,n,m} = \frac{1}{r^{bw}_{k,n,m}} \qquad (22)$$

Here, $r_{k,n}^{cpu}$, $r_{k,n}^{mem}$ and $r_{k,n,m}^{bw} \in [0,1]$ denote the CPU remaining rate (i.e., the amount of residual CPU cores divided by the total amount of CPU cores) on node $n$, memory remaining rate on node $n$ and bandwidth remaining rate on link $(n,m)$.

In Eqs. (22), $c_{cpu k,n}$, $c_{mem k,n}$ and $c_{bw k,n,m}$ increase slowly when the network load is low. On the other hand, if the resource consumption are near to the resource capabilities, $c^{cpu}_{k,n}$, $c^{mem}_{k,n}$ and $c^{bw}_{k,n,m}$ will be very large and increase quickly. Thus, $c^{cpu}_{k,n}$, $c^{mem}_{k,n}$ and $c^{bw}_{k,n,m}$ can be used to indicate the resource bottleneck in the network. Furthermore, we use the sum cost to indicate the load status of a path.

### A. Path Calculation based on Extended Dijkstra's Algorithm

Similar to original Dijkstra's algorithm, the extended Dijkstra's algorithm can return a path from the source node $s$ to every other node on a weighted, directed graph $G_e = (V_e, E_e)$, where $V_e$ is the set of weighted nodes and $E_e$ is the set of weighted edges, each of which is associated with a deployment cost defined above as weight. The algorithm adds the node weight (i.e., deployment cost of edge nodes) besides the edge weight in the topology to obtain the path with least deployment costs. The main idea of this algorithm is that it will adds the node weight to the link weight each time the algorithm detects a node with lower weight. At last, it returns a path with the least deployment cost considering the load status of both edges and edge nodes. By deploying the SFC into the path with the least deployment cost, it can exclude the paths with some bottleneck nodes or edges compared to the original Dijkstra's algorithm.

## B. Resource Allocation Scheme with Minimal Resource Cost

After selecting path, we need to determine each VNF instance's size, i.e., decide the processing resource allocation to each VNF instance. The processing latency of a VNF is decided by the amount of resource demand and resource allocated to it. It is obvious that, to achieve the same latency, a VNF with more resource demand should be allocated more processing resources than a VNF with less resource demand. Fig.1 gives an example that after an available path is selected for SFC $k$, the propagation and transmission latency are known (for the sake of elaboration on processing allocation, we assume the bandwidth allocation is known). Given an E2E service latency requirement, in other words, latency budget, we need to determine the latency distribution on each VNF instance of this SFC $k$. We assume that the total latency budget of SFC $k$ is 10 ms, the communication latency $l_k^c$ on the selected path $P_1$ is 2 ms, thus the remaining latency budget for VNFs are 8 $ms$. We calculate $l_k^p$ for all combinations of CPU core options among all VNF instances and remove the options with exaggeratedly high or low latency, then choose a CPU core allocation option, with least resource cost among the options whose resulted latency is slightly lower than $L_k$. In Fig.1, the CPU core allocation (2 cores, 2 cores, 1 core) for VNF instance 1/2/3 results in the least cost 3.6 $, causing 7.8 $ms$ ≤ 8 $ms$. The algorithm tries to keep the latency experienced by the service within a scope around the latency requirement.

The pseudocode of Det-SFCD is shown in $Algorithm$ 1. For an incoming SFC $k$, we first calculate the deployment costs of CPU, memory and bandwidth according to the current network status in Lines 2-5 and update the weighted topology $G_e$ with the deployment costs in Line 6. With extended Dijkstra's Algorithm, we then derive one available path with least deployment cost in Line 7. In Line 8-9, optimal resource allocation scheme is applied to SFC $k$ and the CPU core allocation to each VNF instances is output to $Q(V_k)$. Then, the algorithm checks if the selected path $P_k$ can satisfy the resource allocation on this SFC $k$ in terms of CPU core, memory and bandwidth. If there are adequate resources along the path SFC $k$, Line 12 determines the embedding of the these VNF instances into the physical nodes along the path $P_k$. The embedding scheme tries to place each VNF instance into the least loaded physical node to keep the load balancing on processing resource. After embedding this SFC $k$, the algorithm updates the network status and preocess the next SFC $k + 1$.

**Algorithm 1: Det-SFCD**

Input: SFC set K
Output: SFC deployment

1. for $k \in$ K do
2.     for $each\ n \in$ N,$(n,m) \in$ E do
3.         The deployment cost on CPU, memory of $n$ and bandwidth of $(n,m)$ at current network status
4.         $c_{k,n}^{cpu}, c_{k,n}^{mem}, c_{k,n,m}^{bw}$ ← Update the weighted topology $G_e$ with and
5.     $P_k$ ← Path with $c_{k,n}^{cpu}, c_{k,n}^{mem}, c_{k,n,m}^{bw}$ least deployment cost
6.     $Q(V_k)$ ← Optimal resource allocation scheme along the selected path $P_k$
7.     $Bool$ ← Check whether $P_k$ satisfies SFC $k$ in terms of CPU $Q(V_k)$, memory and bandwidth allocation
8.     if $Bool == true$ then
9.         Embed the VNF instances (along with virtual links) into the nodes in a load-balancing way
10.        Update the $network\ status$
11.     else 12 Failed

## V. PERFORMANCE EVALUATION

In this section, we demonstrate the performance evaluation of our proposed algorithm. We first present the simulation setup used in the evaluation. Then, we compare our proposed algorithm with the existing benchmark algorithms and evaluate their performance in different cases.

### A. Simulation Setup

We consider a topology [20] (not shown in this paper due to space limitation) with 52 edge nodes, each edge node is associated with 3 CSs. The radio configuration is in line with the RAN VNF parameter specified in [21]. We assume that all SFCs are running with 4 VNFs. The maximum number of CPU cores permitted to be allocated per VNF instance is set as 8. The bandwidth capacity per link is 10 Gbps. The memory capacity and CPU capacity are set as 64 GB and 128 cores, respectively. In the simulation, the source and destination nodes are set randomly. We repeat the simulation in 20 epochs to eliminate contingency, in each of epoch a set of SFCs arrive and leave the environment. The arrival rate of SFC requests follows a tidal distribution and the lifetime of each SFC request obeys the exponential distribution with an average of 100 time units. We set the simulation time as 1000 time units. For each SFC request, the resource block $N_k$, data rate $\lambda_k$(Mbps), and memory(GB) are set as randomly distributed value within [50, 100], [20, 200] and [1,8], respectively [21].

## B. Performance Results

We compare our proposed algorithm with a benchmark algorithm: Shortest Path Heuristic + Latency Equalization (SPH-LE). It first finds the available path by shortest path heuristic without considering deployment cost. Second, it uniformly distributes the latency on each VNF instance and the VNF instances on the path.

As shown in Fig.2(a), the performance of Det-SFCD is evaluated under different latency requirements. It is obvious that our proposed algorithm Det-SFCD outperforms SPH-LE in terms of service acceptance rate. When the network load is high, between [0-400] time units, Det-SFCD achieves much higher service acceptance rate gain compared with the SPHLE algorithm on average 34%. On the other hand, it achieves a lower performance improvement, by an averaging 14%, when the network load is low between 400-600 time units. We can also observe that SFC requests with higher latency requirements need much more resource allocation, which, in turn, results in lower service acceptance rates. Fig.2(b) presents the overall profits of ISP resulting from service revenue and resource cost. Since the acceptance rate of Det-SFCD, shown in Fig.2(b), is higher, which will contribute to the overall profit, we can observe that the proposed Det-SFCD exhibits the better performance in terms of overall profits. On one hand, Det-SFCD accepts more SFC requests by considering the deployment cost to avoid network congestion, which increases the network revenue. On the other hand, by deriving the cost-aware CPU resource allocation for VNF instances under the deterministic latency requirements, Det-SFCD reduces the resource cost for SFC deployment. Finally, we evaluate the performance of Det-SFCD on the mean latency and jitter in Fig.2(c). The results show that the latency experienced by mathematical model of this problem and then propose a novel deterministic SFC deployment algorithm aiming at maximizing the service profits for ISP while ensuring the E2E latency within a deterministic bound. This is by 1) finding an path with the least deployment cost for SFC placement and 2) determining the optimal VNF resource allocation with minimal resource cost for deterministic latency bound. The performance evaluation results show that the proposed Det-SFCD algorithm obtains higher acceptance rate and overall profits compared with the benchmark strategy.

ACKNOWLEDGMENT

This work was partially supported by the European Union Horizon 2020 Research and Innovation Program through the MonB5G Project under Grant No. 871780. It was also supported in part by the Academy of Finland 6Genesis project under Grant No. 318927 and IDEA-MILL with grant number 335936.

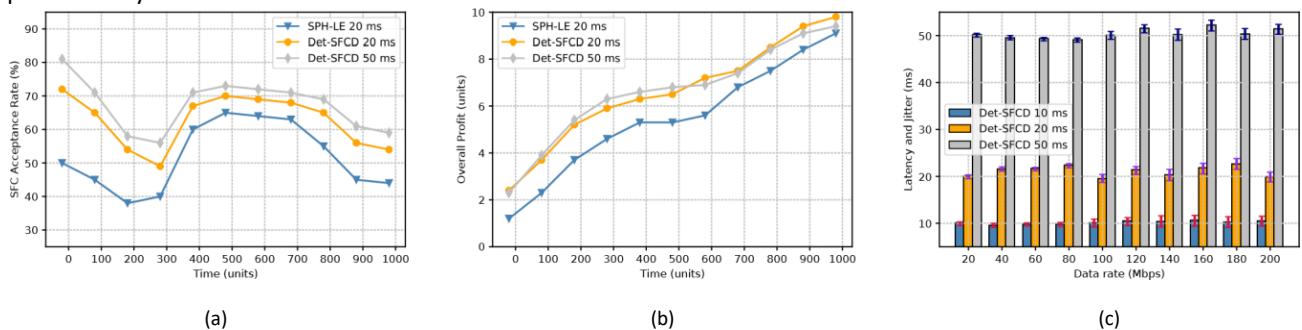

Fig. 2: (a) SFC acceptance rate over time; (b) Overall profit over time; (c) Latency and jitter for different data rates.

SFCs are close to the latency requirement. However, the jitter increases along with the data rates, since the higher required CPU resources allocation will result in higher jitter.

## VI. CONCLUSION

In this paper, we investigated the joint resource allocation and SFC placement problem for deterministic latency assurance in 5G edge networks. We first formulated the